\documentclass[12pt,aps,prb,preprint]{revtex4}   

\usepackage{amsmath}    
\usepackage{graphicx}   

\begin{document}

\title{How numbers help students solve physics problems}
\author{Eugene T. Torigoe}
 \affiliation{Thiel College, Greenville, PA 16125}
 \email{etorigoe@gmail.com}   
\date{\today}

\begin{abstract}

Previous research has found that introductory physics students
perform far better on numeric problems than on otherwise equivalent
symbolic problems.~\cite{Torigoe-06, Torigoe-11}  This paper describes a framework to explain these differences developed by analyzing interviews
with introductory physics students as they worked on analogous numeric and
symbolic problems.  It was found that information about the physical situation, as well as the problem solving process are represented in subtly different ways in numeric problems compared to symbolic problems.  In
almost every respect the inclusion of numbers makes information more
transparent throughout the problem solving process.

\end{abstract}

\maketitle

\section{Introduction}

In introductory physics symbolic algebra is one of the main representations
used to understand physical reality. Unfortunately many students have difficulty solving problems using only symbols.  Introductory physics students are often much more successful on numeric problems than on analogous symbolic problems.  On a
problem studied in an earlier investigation students performed nearly 50
percentage points better on a numeric question than the otherwise
equivalent symbolic problem.~\cite{Torigoe-06}  While
such a large difference is rare, double digit differences in score
are common.~\cite{Torigoe-11}

An analysis of student work showed that many of the errors were consistent with students using symbols in ways that were inconsistent with their definition.  For example, using a symbol defined to be a property of one object as if it were a property of a different object.  These difficulties indicate that many students struggle using and understanding the algebraic equations that are such a vital part of physics problem solving.  

This paper presents a framework which attempts to explain these difficulties in terms of the different ways information is encoded by symbols in numeric and symbolic problems.  From the perspective of this framework numbers help students solve physics problems because numbers make it easy to distinguish symbol states such as the difference between a known and an unknown, and also minimize the possibility of symbol association confusion.  Without numbers, symbols in purely symbolic problems must encode a greater amount of information.  In addition, the interpretation of a symbol may change as the solution progresses.  For example, there is no commonly used notation to denote when a symbol has transformed from an unknown quantity into a known quantity.

The idea of multiple interpretations for an algebraic symbol is not new in the field of mathematics education research.  As part of the Concepts in Secondary Mathematics and Science (CSMS) project studying the mathematical ability of 3000 British school children between the ages of 11 and 16, Kuchemann was able to identify six different interpretations that children applied to letters in algebraic problems.~\cite{Kuchemann-81}

\begin{enumerate}
	\item Letter Evaluated - A letter is assigned a numeric value from the outset
	\item Letter not used - The children ignore the letter
	\item Letter used as an object - The letter is regarded as a shorthand for an object or as an object in its own right.  For example, 2a is interpreted as two apples, in contrast to the interpretation of 2a as 2 times the number of apples.
	\item Letter used as a specific unknown - The children regard a letter as a specific but unknown number, and can operate upon it directly
	\item Letter used as a generalized number - The letter is seen as representing, or at least as being able to take, several values rather than one
	\item Letter used as a variable - The letter is seen as representing a range of unspecified values, and a systematic relationship is seen to exist between two such sets of values.
\end{enumerate}

While the first two categories were improvised methods, the latter four methods reflect legitimate ways that symbols are used in algebra.  Each of those latter four interpretations can be valid depending on the context in which the symbol is introduced.  Similarly in physics, a letter can refer to the unit of a quantity, a specific known or unknown quantity, or as a variable in a general equation.  

Bills suggests that it is the implicit change in the meaning of symbols without explicit notational change that is the source of confusion for many students in algebra.~\cite{Bills-01}  In algebra she identified four common transitions when such implicit changes take place: 1) variable to an unknown to be found, 2) placeholder in a form to an unknown to be found, 3) unknown to be taken as a given to an unknown to be found, and 4) unknown to be taken as given to a variable.

Students often have diffulty making the transition between arithmetic, which primarily deals with numeric computation, and algebra, which has a much greater reliance on symbolic representation.  Kieran studied how prior experience with arithmetic computation skewed students' understanding of the equal sign in algebra.~\cite{Kieran-81, Kieran-90, Kieran-92} She found that many students did not view the equal sign as an equivalence symbol, but rather as a ``do something symbol".  This  interpretation was a generalization of a procedure for solving arithmetic problems, in which the equal sign is commonly associated with a prompt to perform a computation.  When asked to solve problems they sometimes violated the equivalence relation expressed by the equal sign when performing their mathematical actions.  For example, consider the following arithmetic problem.

\begin{quote}
In an existing forest 425 new trees were planted.  A few years later, the 217 oldest trees were cut.  The forest then contains 1063 threes. How many trees were there before the new trees were planted?~\cite{Kieran-81}
\end{quote}

A student might write 1063 + 217 = 1280 - 425 = 855 while solving the problem.  The equal sign is used procedurally by this student as a prompt to evaluate the previous expression, but violates the definition of the equal sign because it is not used as a statement of equivalence between the preceding and following expressions.  Even though the proper understanding of the equal sign was not important for the solution to this particular question, such a misunderstanding could potentially hinder the symbolic representation of relationships required for many algebraic solutions.

Many college students also demonstrate difficulties using and understanding algebraic symbols.  Clement and colleagues studied undergraduate students' ability to mathematically represent relationships between quantities.~\cite{Clement-82}  In these studies he found that many of these students had difficulties coordinating the different meanings of the symbols in the mathematical equations.  He and his colleagues have used variations of what has become known as the ``Students and Professors" problem.

\begin{quote}
Write an equation using the variables S and P to represent the following statement: ``There are six times as many students as professors at this university."  Use S for the number of students and P for the number of professors.
\end{quote}

First-year engineering students scored roughly 60\% on this question, with the most common error being a reversal of the correct quantitative relationship (6S=P instead of S=6P).  While it might first appear that the error was due to carelessness, research involving a variety of techniques demonstrated that many who made the reversal error demonstrated that they conceptually understood the relationship but did not understand how to correctly represent the relationship algebraically.  Many students used the symbols S and P in a way similar to Kuchemann's third category in which the letter is used as shorthand for the object.  For example, 6S was viewed as six students, rather than six times the number of students.  More recently Cohen and Kanim tested student and professor like problems on introductory physics students and found similar results.~\cite{Cohen-05}

Soloway, Lochhead and Clement showed that in the context of writing a computer program that students were much more likely to correctly represent the algebraic relationship.~\cite{Soloway-82}  A class of 100 students was given a students and professors like problem and was split so that half were assigned to write a computer program and half asked to write the algebraic equation.  The students asked to write the computer program performed 24 percentage points better than the group asked to write the algebraic equation.  This finding is consistent with the idea that many students are more comfortable in the context of numeric computation than with symbolic representation.

Trigueros and Ursini performed a study in which they developed a mathematics questionnaire designed to probe first year undergraduates understanding of the different meanings of symbols in algebra.~\cite{Trigueros-03}  The students in the study were 164 first-year undergraduates from a variety of non-science majors who had failed a classification test meant to measure students' preparedness to take calculus.  Each question was created to measure at least one aspect of their understanding of the symbol as an unknown, symbol as a general number, and the symbol in a functional relationship (related variables). They found that the majority of the students in their sample had very poor understanding of these three ways of understanding an algebraic symbol.  The rate of success along each of the three categories was about 50\%.  They concluded that many first-year undergraduate students still primarily hold a non-algebraic perspective on the meaning of symbols.  

Even if students enrolled in introductory physics possess an algebraic perspective and can interpret the different meanings of algebraic symbols, they will find that problem solving in physics is not just the solving of algebra word problems with physical objects. Equations in physics contain a large amount of information, much of which is implicitly encoded and context dependent.  This encoding is so specific to physics that even mathematicians may have difficulty interpreting the meaning of symbolic physics equations.~\cite{Redish-05, Redish-10}  Redish asked both physicists and mathematicians the following question:

\begin{quote}
If $A(x,y) = K(x^2 + y^2)$ with K being a constant, then what is $A(r,\theta)$?
\end{quote}

He reports that while physicists normally recognize the transformation from Cartesian to cylindrical coordinates and answer $A(r,\theta) = Kr^2$, mathematicians commonly view it as a symbol substitution and answer $A(r,\theta) = K(r^2 + \theta^2)$.  For physicists the letter chosen carries information about the type of quantity it is meant to represent.  For them x and y represent the positions along two perpendicular axes, but that is not necessarily the case for mathematicians.  Another example of the difference in interpretation between mathematicians and physicists is the use of the letter x.  In physics the letter x usually represents a distance or position, and therefore cannot be used to represent an unknown as is the case in many mathematics courses.

Sherin has uncovered a context dependent vocabulary used by physicists to interpret symbolic equations.~\cite{Sherin-01}  Sherin documented a series of symbolic forms that describe how equations are interpreted to output conceptual narratives.  In some cases the same equation structure can be interpreted in different ways.  For example, the equation $F_{net} = ma$ is most commonly understood as a causal relationship, whereas the equation $N = mg$ is most commonly understood to represent a balancing of two forces even though both equations have the same mathematical structure: $[ ] = [ ]$.

In most introductory physics courses the range of mathematical abilities of the students is large.  Some students may possess only basic algebraic skills and may mainly rely on arithmetic computational strategies when solving problems.  Others who have a strong understanding of algebra as it is taught in math classes may be confused by the ways symbols are used in physics.  This paper argues that student difficulties with purely symbolic problems are to some extent a result of confusions related to how information is encoded by symbols in physics.  The way information is represented as well as what information is represented is different in numeric problems compared to purely symbolic problems.  Some of the difficulties students have may be a result of inappropriately generalizing methods that work when solving numeric problems but do not when solving symbolic problems.

\section{Problem Solving Interview Protocol}

As an extension of earlier work analyzing numeric and symbolic problems on final exams in a large introductory physics course, speak-aloud problem solving interviews with thirteen introductory physics students was performed.~\cite{subjects}  The goal of these interviews was to uncover mechanisms that explain why numeric problems are often much easier than symbolic problems.  

During the interviews each student was first given the symbolic version to
solve while speaking aloud about their method to reach the solution.
Whether correct or incorrect, the subject was asked questions to
gauge their understanding of the symbols in the problem.  If the
subject had difficulties with the symbolic version, they were asked
to solve the numeric version of the same question. If the subject
was then able to solve the numeric version the subject was then
asked to use their numeric solution to find the correct symbolic
expression. The students were never told whether they found the
correct or incorrect result.  The same procedure was used for all
of the questions.

The choice to give the symbolic version before the numeric version was informed by the earlier results for similar problems in the earlier study involving large student populations.\cite{Torigoe-06}  In that earlier work the numeric versions had an average score between 85-95\%, while the symbolic versions had an average score between 45-60\%.  The expectation was that individual students would be unable to solve the symbolic version, but then later would be able to solve the numeric version of that same problem.

The analysis of the interviews focused on instances when errors were made and on instances when students were incorrect on one version, but correct on the other version.  A framework describing differences in how information was encoded in numeric and symbolic problems was developed to describe the errors observed during the interviews.  A more extensive discussion of the interviews can be found in the author's dissertation.~\cite{Torigoe-08}

\subsection{Subjects}

Thirteen subjects who were students in the calculus-based
introductory mechanics course, Physics 211, at the University of
Illinois were interviewed in spring 2007.  The subjects were chosen
based on their score on the $1^{st}$ exam so that the
interview sample reflected the range of abilities in the student
population.\cite{subjects} To ensure that kinematics was fresh in
their minds, the interviews occurred no later than two weeks after
the $1^{st}$ exam.  The students were paid \$15 and given
two written solutions to past exams in Physics 211 for their
participation.

\subsection{Interview Questions}

\begin{figure}
\begin{center}
\scalebox{0.5}{\includegraphics{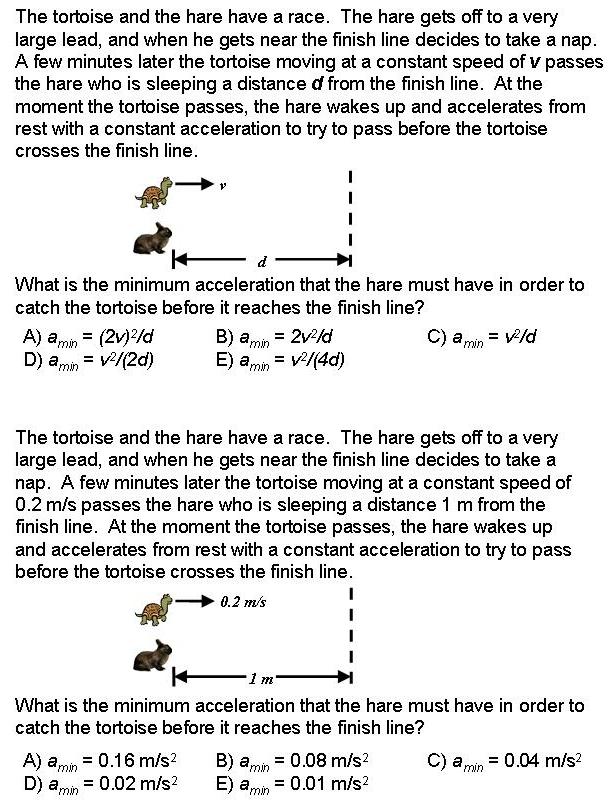}}
\caption{\label{fig:TandH}The numeric and symbolic versions of the
\emph{Tortoise and Hare} question.  This question is analogous to the \emph{Bank
robber} question from the 2006 study.~\cite{Torigoe-06}}
\end{center}
\end{figure}

\begin{figure}
\begin{center}
\scalebox{0.5}{\includegraphics{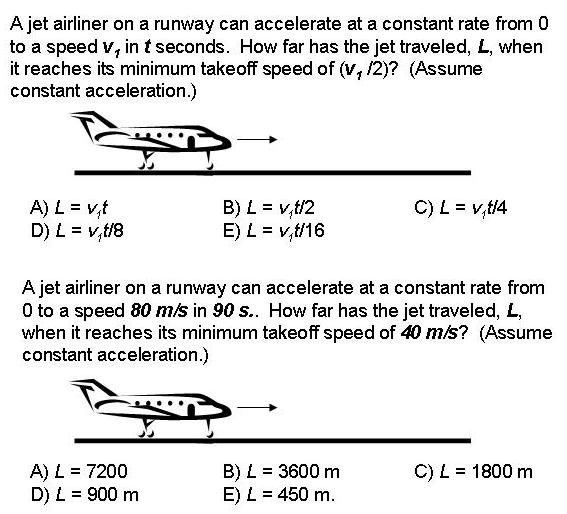}}
\caption{\label{fig:Plane}The numeric and symbolic versions of the
\emph{Airliner} question.  This question is analogous to the \emph{A car can go}
question from the 2006 study.~\cite{Torigoe-06}}
\end{center}
\end{figure}

The physics questions used in this study were modified versions of
the questions used in the earlier final exam study.~\cite{Torigoe-06}  The structure of the questions were the
same but with different surface features.  In the prior study, the performance by a large population of students was used 
to categorize errors, and so the likely errors were known in
advance. The questions used during the
interviews can be seen in Figures \ref{fig:TandH} and
\ref{fig:Plane}.

These questions were designed to contain only simple equation structures to minimize manipulation errors.  While it is likely that manipulating symbolic equations is a difficult for many introductory students, the intent of these interviews was to identify other mechanisms that explain why  students
have difficulty with symbolic problems.  

\section{How symbols encode meaning}

The errors students made, and the problem solving approaches students took were examined and a framework was developed to describe the different types of meaning that are encoded by symbols.  Table~\ref{tab:SymbolStates} lists the different types of information encoded by symbols used in a typical introductory physics problem.  

This framework distinguishes symbol characteristics as either
static or dynamic during the problem solving process. The
term \emph{symbol properties} is used to denote static characteristics such
as object, temporal, or spatial associations.  And the term
\emph{symbol state} to denote whether the symbol is a variable, an
unknown quantity, or a known quantity which can change as the
solution progresses.  

The symbol states category was informed by Kuchemann's categories of the meaning of algebraic letters discussed in the introduction.  The scope of this category was limited to the interpretations commonly used in typical introductory physics problems.  The symbol properties category was an expansion of Redish's description of how letter assignment is used to denote the type of quantity being represented.

\begin{table}
\begin{center}
\begin{tabular}{ll}
\textbf{Symbol States (Dynamic)} & \textbf{Example}\\
\hline\hline 1) General variable & 1) A symbol in a general
equation.  $F$ in
$F=ma$.\\
\hline
2) Unknown & 2) ``Find the force, $F$"\\
\hline
3) Known & 3) ``A force of $F=3N$ pushes ..."\\\\

\textbf{Symbol Properties (Static)} & \textbf{Example}\\
\hline\hline 1) Type of Quantity & 1) Mass, Force, Momentum,
etc.\\
\hline 2) Object Association & 2) Mass of the car, Mass of the
bike.\\
\hline 3) Spatial and Temporal Associations & 3)Initial, final, in,
out.
\end{tabular}
\caption{\label{tab:SymbolStates}Summary of the symbol states for a 
symbol used during physics problem solving, as well as the symbol properties
that are used to specify the relationship of the symbol to the
physical system.}
\end{center}
\end{table}

The problem solving process typical in introductory physics can be described functionally as the
assignment of properties to variables in general equations, and then the transformation of unknown symbols into
known symbols. Symbols begin as variables in general equations,
which rarely carry any specific subject/object, temporal, or axis
associations.  Once the variable is specified with definite
associations it is either a known or unknown symbol. The goal of
most traditional physics questions is the transformation of a target
unknown into a known quantity.  While students rarely confused
symbol states and properties when solving the numeric problems,
confusions were common when working on the symbolic problems.

The ease with which symbol states and properties could be
interpreted during numeric and symbolic problems describe many of
the difficulties observed during the student interviews, which
themselves were consistent with the errors observed in the larger
population. The following sections describe mechanisms
that were uncovered.

\subsection{Symbol Association Errors}

One general category of common errors was to confuse two different
quantities of the same type. For example, using a symbol that is
defined to represent car 1's velocity, as if it were car 2's
velocity.  Both quantities are velocities, but have different object
associations.  This type of confusion was common on
symbolic problems because in purely symbolic solutions two different
quantities of the same type can often be found in the same equation
after equations are combined.  However, in numeric sequential
solutions, where each use of an equation results in a number that
can be plugged into the next equation, this is impossible because
only numeric quantities propagate from one equation to the next.  Thus, the likelihood of such a
confusion in numeric solutions is minimal.

This was a common error when solving the symbolic version of the
Airliner question (shown in Figure~\ref{fig:Plane}). In this problem
an airliner on a runway accelerates at a constant rate from rest.  The airliner's final velocity and the time to reach that
velocity are given, and the question asks you to find the distance traveled when the
airliner reaches half the final velocity.

There were many varieties of errors related to the confusion of two
quantities of the same type. For example, even though the symbol $v$ is defined
as the final velocity, many students used it as if it were half the final velocity.
Similarly the symbol $t$ is defined as the time to reach the final
velocity but was used as the time to reach half the final velocity.
Two interview subjects made this latter error.  These two students
started by using the general equation $x = x_i + v_i t +
\tfrac{1}{2}at^2$. They made the replacements $x_{i} = 0$, $v_{i} =
0$, $x = L$, and $a = v/t$ to get $L = \tfrac{1}{2}(v/t)t^2$.  The
students simplified this to the equation $L = vt/2$. Unfortunately
the symbol $t$ is used to represent two different times.  The $t$ from 
the general equation should properly represent the time to reach a
speed $v/2$, and the $t$ in the acceleration equation should
represent the time the reach a velocity $v$. The students erred when
they canceled the t's after combining the two equations.

Even though they made this error on the symbolic version, both were
able to correctly solve the numeric version. The act of plugging in
numbers seemed to act as a cue to specify the meaning of the
variable they were replacing. Both attempted the same procedure as
they had in the symbolic version, but as they were about to plug in
a value for the time into the kinematics equation, they realized that
it was necessary to solve for the specific time when the jet
airliner reached a speed of 40 m/s (half the final speed), and
that it was inappropriate to cancel the $t$'s as they had in the
symbolic version.

\begin{quote}\emph{
\textbf{Subject F:} Umm, $x= \tfrac{1}{2}a t^2$, and then $a = v/t$
... so when I plugged in the $x$ equals, uhhh, $a = v/t$ in my
equation $x = \tfrac{1}{2}a t^2$, \textbf{I crossed out the times,
but [the t in the acceleration equation] was for when it was 90
[m/s] and [the t in the general equation] is when, we don't know how
long it took}.  So maybe I should...figure out... how
long it takes for the plane to get to 40 m/s...[Subject F then correctly solves the numeric version]\\
\textbf{Interviewer:} OK so how confident do you feel about that?\\
\textbf{Subject F:} Umm, I was pretty confident, but I kind of got
sidestepped over what time I should use, so I went to the side and
solved for it.}\end{quote}

These two students benefitted from the isolation of symbols
with different properties afforded by the numeric procedure.  In this example the inclusion of numbers allowed the students
to isolate each meaning of the symbol $t$ from the other definition
by the use of separate numeric equations.

\subsection{Variable confusion}

The specification of symbol properties is the key step that
differentiates a general equation with variables from an equation
containing symbols with specific properties.

In purely symbolic problems the only way to distinguish a variable
from a specific known or unknown quantity is with a subscript.
Unfortunately, few students in the sample used them.  Some students seemed to treat all symbols in the symbolic version as general variables. 

While there is no special notation for variables in numeric problem
solving, a general equation containing only general variables can be easily identified by the fact that it is purely symbolic, even though numbers are available. When
numbers are plugged into the equation, the remaining symbols transform into 
specific unknown quantities.  These symbols take on the 
symbol properties of the numeric quantities that are plugged into
the equation.  For example, if quantities for the car are plugged
into the equation, then the remaining symbols must all represent unknowns with the 
properties of that car.  

Students may inappropriately carry this interpretation of symbols from numeric problem solving to symbolic problem solving.  There is evidence from the interviews some students believe that the symbols in purely symbolic equations represent general variables.

The failure to distinguish a variable from a specific unknown was
found for some of the subjects to be the cause of the most common
error of the Tortoise and the Hare problem (Figure~\ref{fig:TandH}).
In this problem a Tortoise moving with a constant speed passes a
Hare at rest a certain distance from the finish line.  The instant
the Tortoise passes, the Hare accelerates toward the finish line.
Given the speed of the Tortoise and the distance from the finish
line the question asks you to find the minimum acceleration needed
for the Hare to catch up to the Tortoise before the finish line.

The most common error was to use the equation $v_f^2 = v_i^2 +
2a\Delta x$, to get the incorrect result $a = v^2/(2L)$. This is
incorrect because the symbol $v$ that is given as the
velocity of the Tortoise is used as if it were the velocity of the Hare when
it reaches the finish line.  There were five students in the sample
who made this error.  Those five students were questioned in an attempt to determine if they had a consistent
interpretation of the symbol. Two students said that the symbol represented the velocity of the Tortoise, and three said that it represented the velocity of the Hare.  The two who claimed that the symbol v was the velocity of the Tortoise later demonstrated
a very dynamic interpretation of the symbol.  In the following
exchange, one of those two students switches her interpretation of v in less
than a minute.

\begin{quote}\emph{
[Student selected the answer $a_{min} = v^2/(2L)$]\\
\textbf{Interviewer:} OK and what does $v$ represent? [points to the selected answer]\\
\textbf{Subject A:} umm, the veloc, the \textbf{constant velocity of the tortoise}\\
\textbf{Interviewer:} Alright, so let's say that instead of this
question asking for the minimum acceleration, asked you to find ...
um... the vel, the final velocity of the hare, do you think you
could write down an equation for the final velocity of the hare when
it reaches the finish line?\\
\textbf{Subject A:}  umm, I think I would just, probably rearrange
this equation
[referring to her final answer of $a_{min} = v^2/(2L)$]\\
\textbf{Interviewer:} OK\\
\textbf{Subject A:} Because it, blah, the acceleration does not
change, I mean its constantly accelerating, but in this scenario its
still the minimum acceleration, and distance doesn't
change, so I would  just rearrange the equation to find the final velocity.\\
\textbf{Interviewer:} And that would be the \textbf{final velocity of the hare?}\\
\textbf{Subject A:} \textbf{Right}}\end{quote}

The two students who displayed this behavior seemed to treat the
symbol associations as dynamic properties.  They seemed to falsely
equate the symbols in the final answer as general variables.

For the other three students who held a consistent yet incorrect
definition of the symbol $v$, it is possible that the first time they
considered the meaning of the symbol $v$ was when they were asked
about it.  Because there are no specific cues (like plugging in
numbers) in symbolic problems it is very easy to carry along symbols
without considering their meaning.

\subsection{Confusion of known and unknown symbols}

Many students also exhibited difficulties distinguishing known and
unknown symbols when working on the symbolic versions.  Unlike in mathematics courses where the symbol x in an equation is reserved for unknown quantities, in physics that symbol is already reserved for positions and distances, which can be variable, known, or unknown.  During the
interviews students often lost track of the known and
unknown quantities.  In some cases those confusions led students to inefficient and ineffective
methods.  For example, while solving the Tortoise and Hare symbolic
problem three students started with the equations $d = v_0t$ and $d
= \tfrac{1}{2}a t^2$, and incorrectly eliminated the known quantity
($d$) leaving the two unknowns ($a$ and $t$). As a result they ended
up with one equation and two unknowns.  All three were surprised
when their final result was not one of the answer options.

\begin{quote}
\emph{\textbf{Subject I:} [Starts with equations $d = v_0t$ and $d =
\tfrac{1}{2}a t^2$] um now solving for $a$, I actually solved for,
so $v_ot = \tfrac{1}{2}a t^2$, let's start over, $vt$, $2vt = at^2$
divided $t^2$ [on one side of the equation], divided by $t^2$ [on
the other side of the equation], equals $a$, cross those guys out,
$2v_o/t = a$, and that would give you none of the answers given,
which stinks!... [Answer contains the unknowns $a$ and $t$ because
he eliminated the known quantity $d$]}\end{quote}

These types of errors were not observed in the numeric versions
because known and unknown quantities were easily identified by
either using a number or a letter, respectively.  When solving a symbolic
problem the distinction is implicit because both known and unknown
symbols are represented by letters.  Symbolic questions require that
the solver actively identify the known quantities from the context
in which the symbols are introduced.

\section{Connection to earlier studies}

The mechanisms described in the previous sections are consistent with the results from the earlier study comparing numeric and symbolic problems in a large population introductory physics course.\cite{Torigoe-11}  In that earlier work numeric and symbolic question pairs were analyzed to determine the question properties that were common for problems with large differences between the versions.  The following question properties were common for questions where the symbolic version was significantly more difficult than the numeric version.

\begin{itemize}
	\item \textbf{Multiple equations} - This property distinguishes whether the
problem is commonly solved with one equation or with multiple
equations. 
	\item \textbf{General equation manipulation} - This property signifies whether
it is possible to obtain one of the incorrect choices by combining
general equations or manipulating a single general equation with
minimal changes (for example, replacing $x$ by $d$).
	\item \textbf{Use of a compound expression} - This property signifies that to
reach the correct symbolic solution students must replace a variable
in a general equation with a compound expression (for example,
replacing the variable $v$ by a more specific compound expression
$v/2$).
	\item \textbf{Manipulation error} - This property signifies that a common
error on this question is related to an incorrect manipulation of a
symbolic equation. 
\end{itemize}

The following sections describe the connection between the types of errors identified in the interviews and the results from the prior large population study.

\subsection{Allowing for symbol association errors}

Problems whose solution require multiple equations allows for the possibility of two or more symbols of the same type (two velocities for example) to be present in the same equation when equations are combined.  This does not occur in numeric problems because only numbers propagate from one equation to the next.  If a student solving a purely symbolic problem does not distinguish symbols of the same type with different associations, the student will be susceptible to making an error for problems involving multiple equations.  On the other hand, problems involving only a single equation avoid the possibility of such a confusion.

Students who primarily solve numeric problems may not accustomed to paying attention to the possibility of such errors.

\subsection{Symbols treated as general variables}

In the interviews some students used symbols as if all symbols were general variables with no specific associations.  That observed error is consistent with the properties: general equation manipulation, and use of a compound expression.  It is plausible that if students view symbols as variables that any manipulation of a general equation containing the correct types of symbols would be a reasonable solution.  Similarly, from this perspective the symbol v as a variable could be viewed as a more general version of the term $v/2$.  

\subsection{Poor strategies resulting from symbol state confusion}

\begin{figure}
\begin{center}
\scalebox{0.5}{\includegraphics{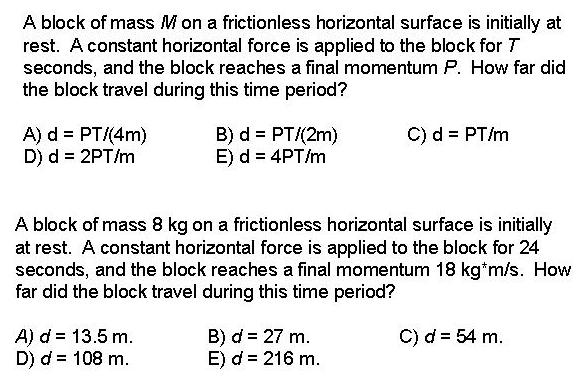}}
\caption{\label{fig:Block}The numeric and symbolic versions of a
question studied to understand how format influences strategic
decisions.}
\end{center}
\end{figure}

While there were no question properties identified in the large population studies that directly related to the difficulty identifying unknowns, it seems plausible that such a confusion could prevent a student from correctly solving a symbolic question.  One possible mechanism for an incorrect result would be that this confusion makes it more difficult to identify the correct strategy to reach the solution.  

An analysis of strategies used to reach the correct answer was undertaken for the two versions of a problem (See Figure ~\ref{fig:Block}) administered in the prior large population study.  The numeric and symbolic versions were given on two versions of the final exam of a
calculus-based introductory physics course.  A total of 765 students saw one of the two versions of this question.  The numeric version had
a average score of 79.6\%. The symbolic version had an average score
of 63.4\%.  The 16.2\% difference between the versions is significant at the $p < 0.001$ level.

The final exam booklets were collected
after the exam, and the students' written work was later analyzed.  A difference in the types of strategies used to reach the
correct answer were observed (See Table~\ref{tab:Strategies}).  To study the differences in strategy this analysis focused only on those students who were able to find the correct final answer.

\begin{table}
\begin{center}
\begin{tabular}{c|c|c}
\textbf{Numeric (N=58)} & \textbf{Symbolic (N=46)} & \textbf{Solution Strategy}\\
\hline\hline

69\% & 28\% & Find velocity with $v = P/m$\\
17\% & 22\% & Find the force with $F = P/t$\\
2\% & 15\% & Start with $a = F/m$\\
0\% & 17\% & Start with $a = v/t$\\
5\% & 11\% & Don't know\\
3\% & 0\% & No Work\\
3\% & 7\% & Other\\

\end{tabular}
\caption{\label{tab:Strategies}The breakdown of strategies used to
find the correct answer on the numeric and symbolic versions.  The
sample analyzed were 58 on the numeric version  and 46 on the
symbolic version. The term ``Find" means that they were able to
solve for an unknown in terms of given quantities. The term ``Start
with" means they began with an equation that contained multiple
unknowns.}
\end{center}
\end{table}

On the numeric version 69\% of students with the correct solution
used the strategy that began by solving for the velocity in terms of
given quantities, $v = P/m$. The students then solved for the
acceleration using the given time, $a = P/(m*t)$. And finally solved
for the distance using $d = \tfrac{1}{2}a t^2$ to find that $d =
Pt/(2m)$. On the symbolic version this strategy was seen in only
28\% of the correct solutions surveyed.  In general, solutions that began with
finding an unknown in terms of known quantities was found in 86\% of
the numeric solutions compared to in only 50\% of the symbolic
solutions.  This result is consistent with the ease of identifying known quantities.

On the other hand, strategies that began with equations with multiple unknowns was seen
in 32\% of the symbolic solutions, compared to only 2\% of numeric
solutions.  These strategies are not very efficient because these
strategies contain at least one step where one set of unknowns is
replaced with another set of unknowns. For example, on the symbolic
version 15\% of the solutions started with the equation $a = F/m$,
which contain the unknowns $a$, and $F$. The students using this
strategy then plugged this into an equation such as $d =
\tfrac{1}{2}at^2$, to get the equation $d = \tfrac{1}{2}(F/m)t^2$,
which contains the unknowns $d$ and $F$. They then used the equation
$F=P/t$ to find $d = Pt/(2m)$.  Even though this leads to the correct result, the replacement of one pair of unknowns for another pair of unknowns is a lateral step that leads no closer to the solution.  Similar
strategies starting with the equation $a = v/t$ were also found.

It is plausible that if difficulties identifying known and unknown quantities in a problem can lead to inefficient strategies that lead to the correct result, then it can also lead to alternative strategies that do not.

\section{Discussion} 

The use of symbols in physics is complex and subtle.  Physics instructors are comfortable using symbols in many different ways, as well as seamlessly transitioning between interpretations.  Unfortunately, many students are not familiar with how symbol meaning changes, and can be confused by their use.  As instructors, we need to be aware of how we use symbols, and also that because of our expertise such transitions between uses can be effortless.  For example, the transition from a variable in a general equation to a quantity with specific associations can occur without a change in notation.  Such transitions should be made explicit to introductory students.

There are also other transitions in symbolic meaning that are common and which often occur without commentary.  One example is the derivation of a general result from a specific case.  In the solution of the specific case, the symbols contain specific associations and are treated as known and unknown quantities.  But in the process of generalizing the result, the symbols in the final equation are transformed into variables with no specific object associations.  

Another example is the use of limiting cases to evaluate a symbolic result.  Again, the solution of the problem involves symbols with specific associations treated as either known or unknown symbols.  When considering limiting behavior the symbols retain their associations, but have variable value.  This combination of symbol properties and symbol states is not seen during the earlier problem solving process.  

The transition in the procedures from numeric to symbolic problem solving is also very subtle.  While every step in a symbolic solution can be mapped to the numeric solution, the way information is represented as well as what information is represented changes from one format to the other.  Students who are comfortable with numeric problem solving may find the procedures they are familiar with do not apply to symbolic problems.  For example, treating the symbols in purely symbolic questions as general variables may work in numeric solutions, but does not in symbolic solutions.  And while tracking the associations of quantities is not important when solving numeric problems, it is very important when solving symbolic problems.

A common practice among physics instructors is to assign numeric
problems, but to instruct the students to solve the problems
symbolically and then only plug in the numbers as one of the last
steps.  While this is a sensible balance between numeric and
symbolic problem solving, one should not assume that the students will be able to easily solve the problem symbolically without instruction or modeling of how to do so.  

One of the main differences between numeric and symbolic problem solving is the necessity to keep track of symbol properties/associations.  Luckily, this can be done easily with subscripts which distinguish symbols with specific properties and
associations from just variables.  In addition, the consideration of
subscripts may act as a cue for students to consider the meaning of
the symbols they are using, just as plugging in numbers did for the 
interview subjects.  Unfortunately, it has been my experience that few students us them without explicit directions to do so.

Another difference is the extra care one must take to identify known and unknown symbols in symbolic problems.  Although students may have implicitly learned from numeric problem solving that
identifying known and unknown quantities is pointless because the
ease of identification in numeric problems, it
is much more helpful when solving symbolic problems.  Some students
may benefit from an explicit notation to distinguish unknowns,
such as circling or underlining the unknown symbols.

For physics experts symbolic representations are very important.
Symbolic solutions allow one to find a general solution, to identify
important variables, and is a method that allows one to check for
errors and sensibility using limiting cases. The importance for
experts makes students' difficulties with symbolic problems all the
more troublesome.  If we expect students to continue on from
introductory physics to a physics major, then it is important that
they become comfortable using symbols.

\begin{acknowledgments}
I would like to thank Gary Gladding, Jose Mestre, Tim Stelzer, Adele' Poynor, Adam Feil, and Michael Scott for commenting on early versions of this paper.  I also would like to thank the members of the physics education research group at the University of Illinois for their input and support over the years.  This material is based upon work supported by NSF DUE 0088734 and NSF DUE 0341261.  
\end{acknowledgments}


\begin{thebibliography}{15}

\bibitem{Torigoe-06}E. Torigoe, G. Gladding, ``Same to Us, Different to Them: Numeric Computation versus Symbolic Representation,"
in 2006 Physics Education Research Conference, edited by L.
McCullough et al.(AIP Press, NY, 2007), pp. 153-156.

\bibitem{Torigoe-11}E. Torigoe, and G. Gladding, ``Connecting
Symbolic Difficulty with Failure in Physics," Amer. J. of Phys.,
{\bf 79(1)}, 133--140 (2011).

\bibitem{Kuchemann-81}D. Kuchemann, ``Algebra," in \emph{Children's Understanding of Mathematics: 11-16}, edited by K. Hart (Murray, London, 1981), pp. 102--119.

\bibitem{Bills-01}L. Bills, ``Shifts in the meanings of literal symbols." in \emph{Proceedings of the 25th conference of the international group for the psychology of mathematics education Volume 2}, edited by M. van den Heuvel-Panhuizen
(Freudenthal Institute, Utrecht, 2001), pp. 161--168. 

\bibitem{Kieran-81}C. Kieran, ``Concepts Associated with the Equality Symbol," Educational Studies in Mathematics. {\bf 12}(3), 317--326
(1981).

\bibitem{Kieran-90}C. Kieran, ``Cognitive processes involved in learning school algebra," in \emph{Mathematics and Cognition: A Research Synthesis by the International
Group for the Psychology of Mathematics Education}, edited by P.
Nesher and K. Kilpatrick (Cambridge University Press, Cambridge,
1990), pp. 96--112.

\bibitem{Kieran-92}C. Kieran, ``The learning and teaching of school algebra," in \emph{Handbook of Research on Mathematics Learning and Teaching}, edited by D. Grouws (MacMillan, New York, NY, 1992), pp. 390--419.

\bibitem{Clement-82}J. Clement, ``Algebra word problem solutions: Thought processes underlying
a common misconception,'' J. Res. Math. Educ. {\bf 13}(1), 16--30
(1982).

\bibitem{Cohen-05}E. Cohen and S. E. Kanim, ``Factors influencing the algebra ``reversal error'',''
Am. J. Phys. {\bf 73}(11), 1072--1078 (2005).

\bibitem{Soloway-82}E. Soloway, J. Lochhead, and J. Clement, ``Does computer programming enhance problem solving ability? Some positive evidence on algebra word problems," in \emph{Computer Literacy},
edited by R. J. Seidel, R. E. Anderson, and B. Hunter (Academic
Press, Burlington, 1982), pp. 171--201.

\bibitem{Trigueros-03}M. Trigueros, and S. Ursini, ``First-year Undergraduates' Difficulties in Working with Different Uses of Variable," in \emph{CBMS Issues in Mathematics Education Volume 12},
edited by A. Seldon, E. Dubinsky, G. Harel, and F. Hitt (American Mathematical Society, Providence, 2003), pp. 1--28.

\bibitem{Redish-05}E.F. Redish, ``Problem Solving and the Use of Math in Physics Courses," Proc.
World View on Physics Education in 2005: Focusing on Change (Delhi, 2005).

\bibitem{Redish-10}E.F. Redish, and A. Gupta, ``Making Meaning with Maths in Physics: A Semantic Analysis," in GIREP-EPEC and PHEC 2009 INTERNATIONAL CONFERENCE August 17-21, University of Leicester, UK, Edited by D. Raine, C. Hurkett and L. Rogers, (The Centre for Interdisciplinary Science, University of Leicester, 2010), pp. 244--260.

\bibitem{Sherin-01}B. Sherin, ``How Students Understand Physics Equations,"
Cog. and Instr., {\bf 19(4)}, 479--541 (2001).

\bibitem{Torigoe-08}E. Torigoe, What Kind of Math Matters? A Study of the Relationship Between Mathematical Ability and Success in Physics, Ph. D. Dissertation,
University of Illinois at Urbana-Champaign, 2008.

\bibitem{subjects}To protect the identity of the interview subjects, each subject in the following transcript excerpts are identified by letters. 


\end{thebibliography}
\end{document}